\documentclass[12pt]{article} 
\usepackage{graphicx}
\begin{document}
\newcommand{\AM}{\ensuremath{\mathcal{P}}}
\newcommand{\CPgf}{\ensuremath{\mathcal{C}}}
\newcommand{\COPgf}{\ensuremath{\mathcal{C}^{\rm O}}}
\newcommand{\CNAPgf}{\ensuremath{\mathcal{C}^{\rm NA}}}
\newcommand{\ACPgf}{\ensuremath{\mathcal{C}_1}}
\newcommand{\ACOPgf}{\ensuremath{\mathcal{C}_1^{\rm O}}}
\newcommand{\ACNAPgf}{\ensuremath{\mathcal{C}_1^{\rm NA}}}
\newcommand{\RG}{\ensuremath{\mathcal{R}}}

\title{Osculating and neighbour-avoiding polygons on the square lattice}
\author{Iwan Jensen\thanks{e-mail: I.Jensen@ms.unimelb.edu.au} \\
Department of Mathematics \& Statistics, \\
The University of Melbourne, Victoria 3010, Australia}
\date{\today}
\maketitle
\bibliographystyle{plain}
\begin{abstract}
We study two simple modifications of self-avoiding polygons. 
Osculating polygons are a super-set in which we allow the perimeter of the 
polygon to touch at a vertex. Neighbour-avoiding polygons are only allowed to 
have nearest neighbour vertices provided these are joined by the associated
edge and thus form a sub-set of 
self-avoiding polygons. We use the finite lattice method to count the number 
of osculating polygons and neighbour-avoiding polygons on the square lattice. 
We also calculate their radius of gyration and the first  
area-weighted moment. Analysis of the series confirms exact predictions for
the critical exponents and the universality of various amplitude combinations. 
For both cases we have found exact solutions for 
the number of convex and almost-convex polygons.
\end{abstract}

\section{Introduction}

A self-avoiding polygon (SAP) can be defined as a walk on a lattice which
returns to the origin and has no other self-intersections. Generally SAPs are 
considered distinct up to a translation, so if there are $p_n$ SAPs of length
$n$ there are $2np_n$ walks (the factor of two arising since the walk can go 
in either direction). The enumeration of self-avoiding polygons on various
lattices is an interesting combinatorial problem in its own right, and is also
of considerable importance in the statistical mechanics of lattice models
\cite{Hug}. Despite strenuous effort this problem has not been solved on any
regular two dimensional lattice. However, progress has been made in the study 
of restricted classes of polygons and many problems have been solved exactly. 
These include staircase polygons \cite{Temperley,Polya,Delest,Brak90,Lin91}, 
convex polygons \cite{Delest,Kim,GE88b,Lin88}, row-convex polygons 
\cite{Brak90,Lin91}, and almost convex polygons \cite{Lin92}. 

\begin{figure}
\begin{center}
\includegraphics{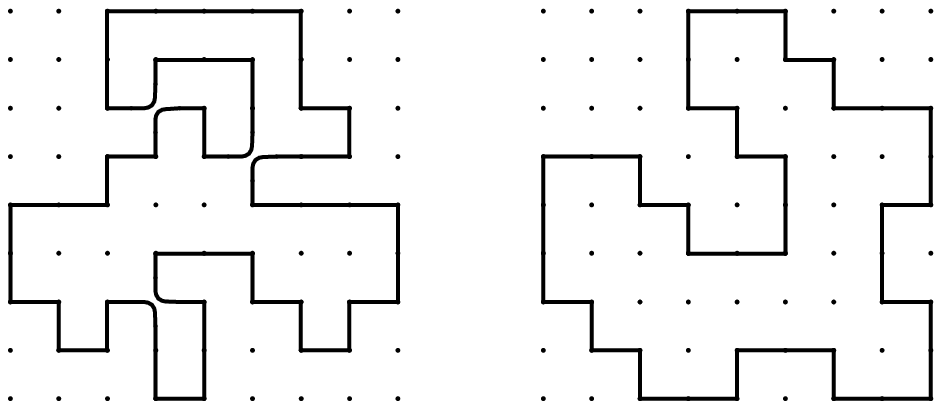}
\end{center}
\caption{\label{fig:onap} 
An example of an osculating polygon (left panel) and
a neighbour-avoiding polygon (right panel).}
\end{figure}

In this paper we study two simple modifications of SAP. Osculating polygons
(OP) form a super-set in which we allow the walk to touch at a vertex but not 
to cross. Neighbour-avoiding polygons (NAP) are only allowed to have 
nearest neighbour vertices when the adjoining edge is part of the perimeter
of the polygon (any two occupied vertices are separated by at least one 
empty vertex) and they are a sub-set of SAP. See figure~\ref{fig:onap} 
for an example of each case. Osculating polygons were introduced in
\cite{JG98} while neighbour-avoiding polygons were studied in \cite{BW1}
as the special limiting case of interacting polygons with infinitely strong
repulsion. We use the finite lattice method to count the
number of osculating and neighbour-avoiding polygons on the square lattice. 
We also calculate their radius of gyration and the first and second 
area-weighted moments. 
The quantities we consider are: (i) the polygon generating
function, $\AM (u)= \sum p_n u^n$; (ii) $k^{th}$ area-weighted moments 
of polygons of perimeter $n$, $\langle a^k \rangle _n$; 
and (iii) the mean-square radius of gyration of polygons of perimeter $n$, 
$\langle R^2 \rangle _n$. These quantities are expected to behave as

\begin{eqnarray}\label{eq:coefgrowth}
p_n & = & B\mu^n n^{\alpha-3}[1+o(1)], \nonumber \\
\langle a^k \rangle _n & = & E^{(k)}n^{2k\nu}[1+o(1)], \\
\langle R^2 \rangle _n & = & D n^{2\nu}[1+o(1)], \nonumber
\end{eqnarray}
\noindent
where $\mu = u_c^{-1}$ is the reciprocal of the critical point of the
generating function, and $\alpha =1/2$ and $\nu = 3/4$ are known
exactly \cite{Nienhuis}, though non-rigorously, in the case of the
honeycomb lattice. Very firm evidence exists from
previous numerical work that the exponent $\alpha$ is universal and thus 
equals 1/2 for all two dimensional lattices \cite{GE88a,EG89,JG98}. The value
$\nu=3/4$ is likewise  well supported by the existing numerical evidence
\cite{PR,GE88a,EG90,Jensen00}. It is also known \cite{CG} 
that the SAP amplitude combination $E^{(1)}/D$ is universal, and that

\begin{equation}\label{eq:BDampl}
BD = \frac{5}{32 \pi ^2}\sigma a_0,
\end{equation}
\noindent
where $a_0$ is the area per site and $\sigma$ is an integer such that
$p_n$ is non-zero only if $n$ is divisible by $\sigma$. For the 
square lattice $a_0=1$ and $\sigma=2$. These predictions have
been confirmed numerically for SAP on many different lattices 
\cite{CG,Lin98,Jensen00}. We would expect the universality of
these amplitude combinations to hold for NAP. What happens for OP
is not immediately clear. The derivation in  \cite{CG} relies on
the self-avoidance and indeed the universality does not extend to
trailgons (trails are walks which are allowed to share a vertex but {\em not}
an edge) which are a super-set of OP. We shall provide compelling evidence
that with a suitable definition of the radius of gyration the universality
does extend to OP.

Any polygon has a minimal bounding rectangle. A polygon is said to be convex if 
its perimeter has the same length as its minimal bounding rectangle. More 
generally one can classify polygons according to their concavity $c$, with the 
perimeter of a polygon of concavity $c$ exceeding the length of the minimal
bounding rectangle by $2c$. For both OP and NAP we have found exact solutions 
for the generating functions for convex and almost-convex ($c=1$) polygons.

In the next section we briefly describe the generalisation of the finite 
lattice method required in order to enumerate OP and NAP. The results of
the analysis of the series are presented in Section~\ref{sec:analysis}.
Exact results for the generating functions for convex and almost-convex 
OP and NAP are presented in Section~\ref{sec:convex}.

\section{Enumeration of polygons \label{sec:flm}}

The method used to enumerate OP and NAP on the square lattice are simple
generalisations of the method devised by Enting \cite{Ent} in his 
pioneering work and includes the significant enhancements employed in previous
work \cite{JG99,Jensen00}. The general method has been described in some detail 
in these papers and for this reason we shall be very brief and only give the 
absolute essential information.

\begin{figure}
\begin{center}
\includegraphics[scale=0.6]{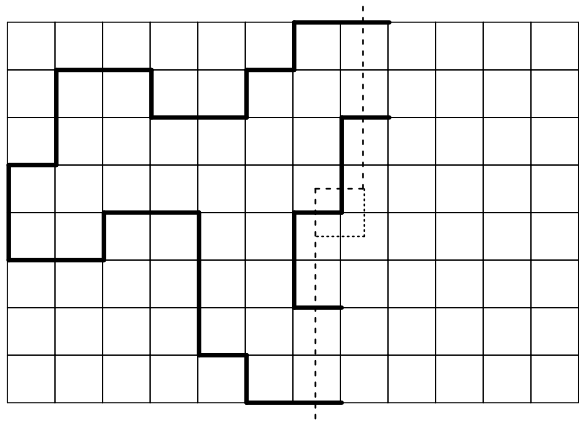}
\end{center}
\caption{\label{fig:transfer} 
A snapshot of the boundary line during the transfer-matrix calculation
for polygons.
}
\end{figure}

The first terms in the series for the polygon generating function can be 
calculated using transfer matrix techniques to count the number of polygons in 
rectangles $W+1$ edges wide and $L+1$ edges long. The transfer matrix technique  
involves drawing a boundary through the rectangle intersecting a set of $W+2$ 
edges. For each configuration of occupied or empty edges along the boundary we 
maintain a (perimeter) generating function for partially completed polygons. 
Polygons in a given rectangle are enumerated by moving the boundary so as to 
add one site at a time as shown in figure~\ref{fig:transfer}. Configurations are 
represented by a set of edge states $\{n_i\}$, where

\begin{equation}\label{eq:states}
n_i  = \left\{ \begin{array}{rl}
 0 &\;\;\; \mbox{empty edge},  \\ 
1 &\;\;\; \mbox{lower part of loop}, \\
2 &\;\;\; \mbox{upper part of loop}. \\
\end{array} \right.
\end{equation}
\noindent
Reading configurations from the bottom edge up, the partially completed
polygon in figure~\ref{fig:transfer} is encoded as $\{1010210202\}$
before the move of the boundary line and $\{1010000202\}$ after the move.
The rules for updating the partial generating functions are described in 
\cite{Ent} and we refer the interested reader to this paper for
further details.

In the following sections we shall briefly outline the changes required in
order to enumerate OP and NAP.

\subsection{Osculating polygons}

\begin{figure}
\begin{center}
\includegraphics{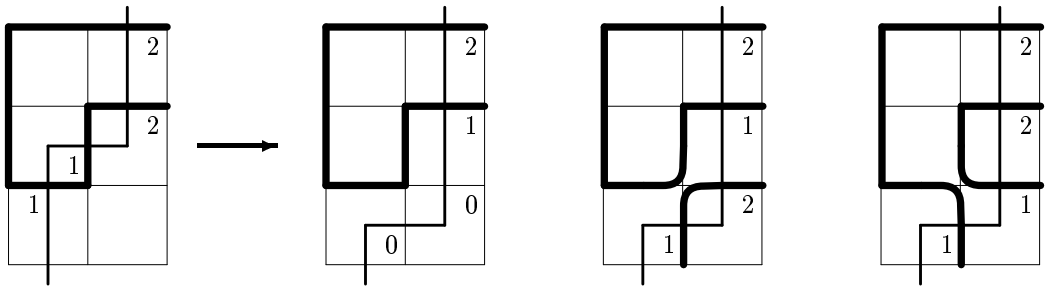}
\end{center}
\caption{\label{fig:OPupdate} 
The input state $\{ 01122\}$ and the three possible output states,  $\{ 00012\}$
where the lower ends are joined with no further action,  $\{ 01212\}$
where the lower ends are joined and a new loop is inserted, and  $\{ 01122\}$
where the lower ends `bounce' of one-another. 
}
\end{figure}

OP are allowed to have vertices on which the perimeter touch but no crossing
takes place. Obviously such vertices are of degree 4 and in terms of the
updating rules are vertices with two incoming and outgoing occupied edges.
The possibilities are thus that the incoming edges are in states $\{11\}$, 
$\{22\}$, $\{21\}$, or $\{12\}$. The first three of these are essentially 
identical in that they permit similar outputs. From $\{11\}$ we can produce the 
outputs $\{00\}$ (with appropriate relabelling of the remainder of the 
configuration as per the original SAP problem) and the two additional outputs 
$\{11\}$ and $\{12\}$. The three possibilities are illustrated in 
figure~\ref{fig:OPupdate} and correspond to cases where, respectively, the 
edges are joined with no further action, bounce of one-another, or are joined 
and a new loop is inserted. The case $\{12\}$ is special in that joining the 
edges results in a closed loop and thus the formation of a separate component. 
This is only allowed provided there are no other occupied edges and the result 
is a valid polygon. This leaves us just with the `bouncing' output $\{12\}$ as 
illustrated in figure~\ref{fig:OPupdate}. The updating rules for the 
transfer-matrix algorithm are summarised in Table~\ref{tab:OPupdate}.

At this point it is pertinent to give some detail about our calculation
of the radius of gyration. The general algorithm was described in 
\cite{Jensen00} but we made some changes to accommodate the degree 4
vertices of OP. In the basic definition \cite{PR} one calculates  the 
radius of gyration of the vertices contained in the perimeter of the
polygon. The problem is how to count the degree 4 vertices of OP. We have 
chosen to rely on the definition of polygons as walks which return to the
starting point. Viewed in this light it is natural to count degree
4 vertices twice, once for each time the walk visits the vertex. This 
makes the necessary generalisation of the algorithm quite simple
and no further details will be given. As we shall see when analysing
the series this definition also ensures that the amplitude combinations
$BD$ and $E^{(1)}/D$ remain universal. 

We calculated the number of OP with perimeter up to $n=90$ and their
radius of gyration and area-weighted moment up to $n=82$.

\subsection{Neighbour-avoiding polygons}

The generalisations to NAP is a little more involved, but the required 
generalisation has been described in some detail in \cite{BW1} and we shall 
thus be brief. In order to ensure the nearest neighbour avoidance we have 
to encode some extra information in the boundary configuration. Essentially
we introduce an extra state `3', which labels an empty edge along which
a contact could occur, i.e. an empty edge next to the perimeter of
the polygon. As an illustration consider again  the partially completed
polygon in figure~\ref{fig:transfer}. This is now encoded as $\{1013213202\}$
before the move of the boundary line and $\{1013333202\}$ after the move.
We are not allowed to occupy any edge adjacent to an edge in state `3' since 
this would result in an illegal contact. This restriction means the updating 
rules can depend not only on the edges in the kink itself (as for SAP and OP)
but also on the states of nearby edges.  The rules for allowed transition of the 
boundary line in the transfer-matrix algorithm are summarised in 
Table~\ref{tab:NAPupdate}.

We calculated the number of NAP with perimeter up to $n=86$ and their
radius of gyration and area-weighted moment up to $n=82$.

\section{Analysis of series \label{sec:analysis}}

The series have exponentially growing coefficients, with sub-dominant term 
given by a critical exponent. The generic  behaviour is 
$G(u) =\sum_n g_n u^n \sim (1-u/u_c)^{-\xi},$ 
and hence the coefficients of the generating function 
$g_n \sim \mu^n n^{\xi-1}$, where $\mu = 1/u_c$. To obtain the singularity 
structure of the generating functions we used the numerical method of 
differential approximants \cite{Guttmann89}. Combining the relationship given 
above between the coefficients in a series and the critical behaviour of 
the corresponding generating function with the expected behaviour 
Eq.~(\ref{eq:coefgrowth}) of the mean-square radius of gyration and moments 
of area yields the following prediction for their generating functions:

\begin{eqnarray}\label{eq:genfunc}
\RG^2_g (u)& = &\sum_n p_n\langle R^2 \rangle _n n^2 u^n =
    \sum_{n} r_n u^n \sim R(u)(1-u/u_c)^{-(\alpha+2\nu)}, \\
\AM^{(k)} (u)& = &\sum_n p_n \langle a^k \rangle _n u^n =
    \sum_{n} a^{(k)}_n u^n \sim a^{(k)}(u)(1-u/u_c)^{2-(\alpha+2k\nu)},
\end{eqnarray}
\noindent
As stated previously the exponent $\alpha=1/2$ while $\nu=3/4$.   
Since the series only contain even terms and the smallest polygon has
size 4, we actually analyse the series $P(x) = \sum_n p_{2n+4} x^n$,
and so on, and obtain estimates for $x_c=u_c^2=1/\mu^2$.
 
Our use of differential approximants have been described in detail in
\cite{JG99,Jensen00} so suffice to say that estimates for the critical
points and exponents are obtained by averaging over many separate 
approximants each using most of the series coefficients and yielding
individual estimates. A very rough and not necessarily reliable error estimate
is obtained from the spread among the approximants. 

Next we turn our attention to the ``fine structure'' of the asymptotic form 
of the coefficients. In particular we are interested in obtaining accurate
estimates for the amplitudes $B$, $D$ and $E^{(1)}$. We do this
by fitting the coefficients to the assumed form (\ref{eq:coefgrowth}).
The asymptotic form of the coefficients $p_n$ of the SAP generating 
function was studied in detail previously \cite{CG96,JG99}.
As argued in \cite{CG96} there is no sign of non-analytic 
corrections-to-scaling exponents to the polygon generating function 
and one therefore finds that
\begin{equation}\label{eq:polasymp}
p_n = \mu^n n^{-5/2} \sum_{k=0} a_k/n^k.
\end{equation}
\noindent
This form was confirmed with great accuracy in \cite{JG99}. For the
radius of gyration coefficients we found in \cite{Jensen00} that
we had to take direct account of the correction-to-scaling exponent
$\Delta=3/2$ which leads to the  asymptotic form

\begin{equation}\label{eq:rgsasymp}
r_n = \mu^n n [BD + \sum_{k=0} a_k/n^{k/2}].
\end{equation}
\noindent
Alternative we could also fit to the form
\begin{equation}\label{eq:rgsratasymp}
r_n/p_n = n^{7/2} [D + n^{5/2}\sum_{k=0} a_k/n^{k/2}].
\end{equation}
\noindent
Asymptotic forms similar to those above also hold for the area-moment
coefficients after the appropriate change of the leading exponent.

\subsection{Osculating polygons}

\begin{figure}
\begin{center}
\includegraphics[scale=0.95]{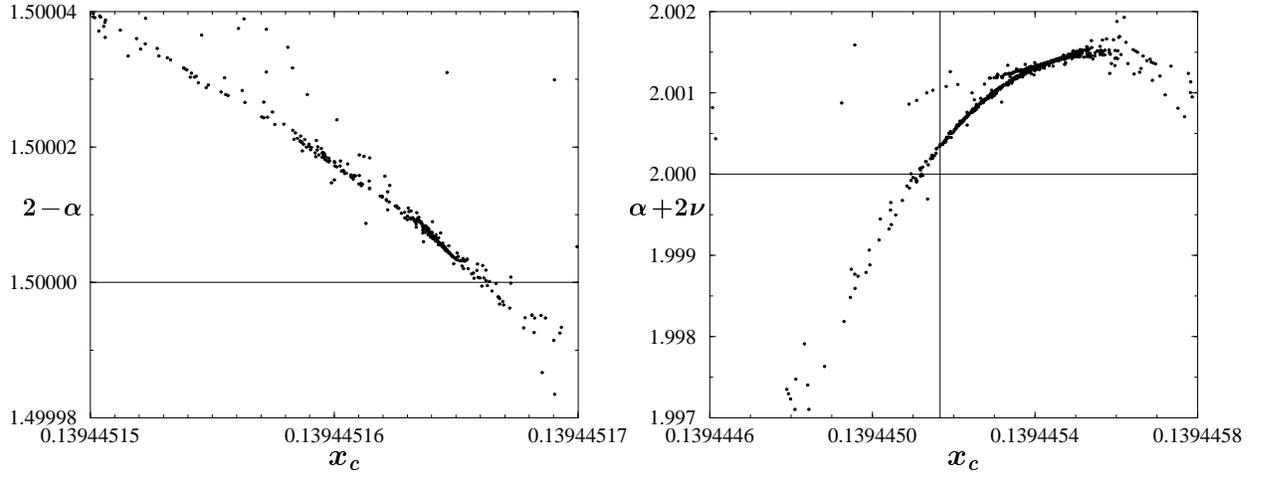}
\end{center}
\caption{\label{fig:OPcrit} 
Estimates for the critical exponent $2-\alpha$ (left panel) and
$\alpha+2\nu$ (right panel) vs. the corresponding estimates for critical 
point $x_c$ as obtained from third order differential approximants to the 
OP generating function and radius of gyration series. The horizontal lines
indicate the expected exponent values while the vertical line in the right
panel indicates the estimate for $x_c$ obtained from the data in the left panel.
}
\end{figure}

In Table~\ref{tab:OPcrit} we have listed estimates for the critical point and 
exponents obtained from second and third order differential approximants to the 
polygon generating function, radius of gyration, and first area moment series.
It is obvious that the OP generating function has an exponent consistent
with the exact values $2-\alpha=3/2$, as we would expect. It is equally
clear that the first moment as expected has a logarithmic singularity
thus confirming $\nu=3/4$. From the polygon series we also find an accurate 
estimate for the critical point $x_c=0.139445164(3)$. The estimates for the
exponent of the radius of gyration series differs slightly from the expected
value of $-(\alpha + 2\nu)=-2$. However, the estimates obtained from this series 
are also much less accurate than those from the other series and the value $-2$
can certainly not be ruled out. In order to try to obtain a more accurate
estimate of $x_c$ and resolve the discrepancy of the radius of gyration
series we look in greater detail at estimates from the individual differential
approximants. In figure~\ref{fig:OPcrit} we have plotted estimates of the
critical exponents vs. the corresponding critical points. From the data
in the left panel we see an essentially linear relationship between estimates for
the exponent $2-\alpha$ and $x_c$, We notice that the exponent attains
its expected value 3/2 when $x_c = 0.1394451660(5)$, which we take as our
final estimate for $x_c$, and we thus find $\mu^{\rm O}=2.677924128(5)$. 
As one would expect $\mu^{\rm O}$ is larger than the connective constant
$\mu=2.63815853034(10)$ for SAP.
From the data for the radius of gyration we
notice that the curve traced by estimates comes very close to crossing through
the point given by the expected exponent value and the value for $x_c$ just
obtained. We also note that in a similar plot using second order approximants
the curve was further removed from this point. We take this as firm evidence
that given a sufficiently long series and high enough order approximants
any discrepancy between the behaviour of the actual estimates and the
expected  behaviour would be completely resolved.

\begin{figure}
\begin{center}
\includegraphics[scale=0.80]{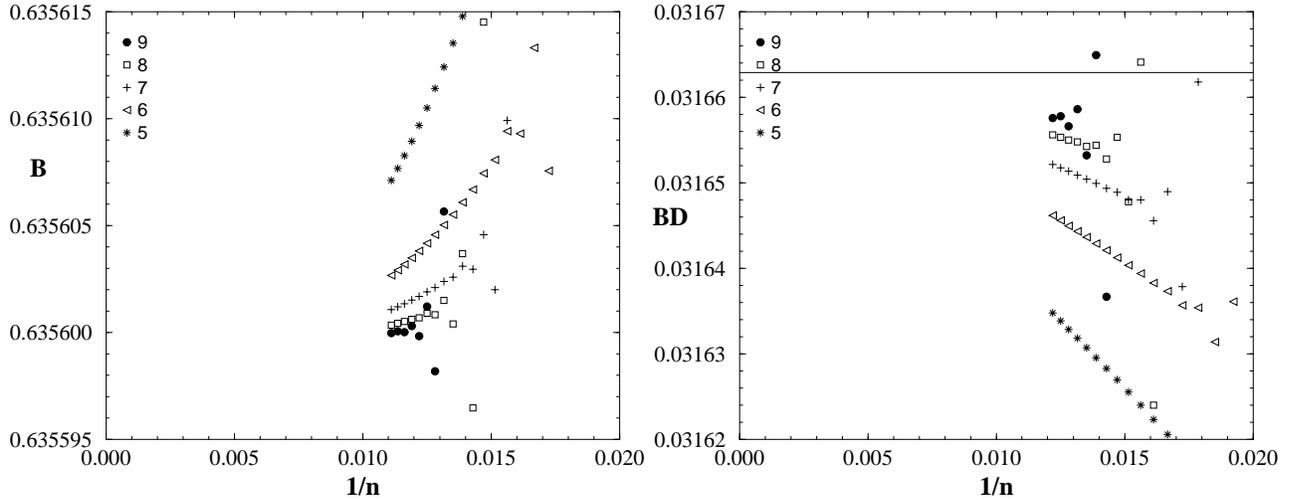}
\end{center}
\caption{\label{fig:OPampl} 
Estimates for the leading amplitudes $B$ (left panel) of the OP generating function
and amplitude combination $BD$ (right panel) from the radius of gyration series 
vs. $1/n$. In each case we show data using varying numbers of terms in the 
expansions in Eq.~(\protect{\ref{eq:polasymp}}) and (\protect{\ref{eq:rgsasymp}}),
respectively. The horizontal line in the second panel indicate the expected exact
value of the amplitude combination $BD$.  
}
\end{figure}

Having confirmed that the exponents have their expected values and obtained
a very accurate estimate for $x_c$, we turn our attention to the leading
amplitudes. First we note that the expected asymptotic form given in
Eq.~(\ref{eq:polasymp}) is completely confirmed by repeating the analysis
carried for the SAP case \cite{JG99}.  The results for the leading amplitude 
are displayed in figure~\ref{fig:OPampl}. We notice that all fits appear to 
converge to the same value as $n \rightarrow \infty$, and that, as more and 
more correction terms are added to the fits the estimates exhibits
less curvature and that the slope become smaller (although the fits using 
9 terms are a little inconclusive).  This is very strong evidence that 
Eq.~(\ref{eq:polasymp})  indeed is the correct asymptotic form of $p_n$. We 
estimate that $B=0.6355995(10)$. The corresponding plots for the amplitude
combination $BD$, obtained from the radius of gyration series, clearly
show that these estimates are consistent with the exact value
$BD=5/16\pi^2$, and we thus estimate $D=0.04981575(15)$. 
A similar analysis of the area-moment series yields the estimate
$E^{(1)}=0.12520(1)$. From this we find the amplitude ratio
$E^{(1)}/D=2.51326(20)$ in remarkable agreement with the estimate 
$E^{(1)}/D=2.51326(3)$ from the SAP series \cite{Jensen00}.

\subsection{Neighbour-avoiding polygons}

Table~\ref{tab:NAPcrit} lists estimates for the critical point and exponents 
for the NAP series. Albeit the accuracy of these estimates is somewhat
poorer than before there can be no doubt that the exponent estimates
are consistent with the expected exact values. Again we use a plot of 
$2-\alpha$ vs. $x_c$  to obtain the improved estimate $x_c=0.1864580(5)$ 
and thus $\mu^{\rm NA}=2.315845(4)$. As expected this connective constant
is smaller than the SAP counter-part. We note that the
corresponding value for the walk generating function $u_c=\sqrt {x_c} =
0.4318078(6)$ is in full agreement with the unbiased estimate $u_c=0.43180(2)$ 
given in \cite{BW1}, but quite a bit larger than the same papers biased estimate
$u_c=0.4317925(1)$. The only possible explanation we can think off for this
discrepancy is that perhaps the exponent used in biasing was wrong.
Cardy \cite{Cardy} predicted that excluding walks with parallel nearest-neighbour 
steps should result in a change in the exponent $\gamma$ of the walk generating 
function.  Numerical studies confirmed a slight change in exponent for Manhattan 
lattice SAW \cite{BW2} (by it very definition the Manhattan lattice does not
allow any parallel nearest-neighbour steps). Clearly neighbour-avoiding walks
have no nearest-neighbour steps at all. In \cite{BW1} the authors used the
usual SAW exponent $\gamma=43/32$, which thus could be the wrong choice for
neighbour-avoiding walks. In fact one might see the discrepancy with the
biased estimate for $u_c$ as further evidence that there is a change
in the exponent $\gamma$ for walks with no parallel nearest-neighbour steps.

For NAP the analysis of the radius of gyration and area-moment series confirm 
the universality of $BD$ and $E^{(1)}/D$. Using this and the direct analysis
of the polygon series we arrive at the amplitude estimates $B=0.36955(5)$,
$D=0.08568(1)$ and $E^{(1)}=0.21534(2)$

\section{Exact results for convex and almost-convex polygons \label{sec:convex}}

In section~\ref{sec:flm} we described how the series expansion for the polygon 
generating function was calculated by counting the number of polygons spanning
various $W \times L$ rectangles. From this data it is obviously easy to extract
the corresponding series for the generating functions for convex polygons with
concavity $c$. In the following we use these series to calculate the exact 
generating functions for convex and almost-convex OP and NAP.

Throughout this section we use the variable $x$ as the conjugate of the
semi-perimeter of the polygons. In other words the coefficients of
$x^k$ is the number of polygons of perimeter $n=2k$.

\subsection{Convex polygons}

The generating function for convex osculating polygons (COP) is quite easy
to derive. First of all we note that any ordinary convex polygon is a COP.
So all we need calculate is the generating function for polygons with at least
one osculation. It is not difficult to see that these polygons can be obtained
by simple concatenations. In the simplest case we can take a convex polygon 
whose top right-hand corner is also the corner of the minimal bounding rectangle 
and concatenate with a convex polygon whose bottom left-hand coincides with 
the bounding rectangle (obviously these two types of polygons have identical 
generating functions). Due to to the convexity constraint the only way to
get more osculations is by inserting convex polygons where {\em both} the
top right-hand and bottom left-hand corners are shared with the minimal bounding
rectangle. Thankfully the generating functions for these types of polygons
are already known. The generating function for the first type of polygons
was calculated by Lin and Chang \cite{Lin88} and in their honour we shall
stick with their notation and call this function $H(x)$. The second 
intermediate type of convex polygons are the well-known and much loved
staircase polygons, their generating function is $S(x)$. Since we can
insert any number of staircase polygons and go along either diagonal of
the square lattice we find that the generating function, $\COPgf$, for convex
osculating polygons is

\begin{equation}\label{eq:cop}
\COPgf (x) = \CPgf (x) + 2H(x)/(1-S(x)),
\end{equation}
\noindent
where $\CPgf$ is the generating function for convex polygons.

After inserting the known generating functions 

\begin{eqnarray*}
\CPgf (x) & = & x^2(1-6x+11x^2-4x^3)(1-4x)^{-2}-4x^4(1-4x)^{-3/2}, \\
H(x) & = & x^2(1-4x)^{-1/2}, \\
S(x) & = & (1-2x-\sqrt{1-4x})/2,
\end{eqnarray*}
\noindent
and doing a little arithmetic we find

\begin{equation}\label{eq:copexp}
\COPgf (x) =x^2(2-10x+14x^2-5x^3-4x^4)/[(1-4x)^2(2+x)]
-x^3(1+2x)^2/[(1-4x)^{-3/2}(2+x)].
\end{equation}

The generating function, $\CNAPgf$, for convex neighbour-avoiding polygons (CNAP)
is simply $\CNAPgf (x) =x^2[1+ \COPgf (x)]$. In other words, with the exception
of the simplest polygon of size $n=4$, the number of COP is
identical to the number of CNAP with four more steps. This means that there
should be a simple bijection between COP with perimeter $n$ and CNAP with
perimeter $n+4$. To describe this bijection we need a slightly different
way to describe a convex polygon. We can define convex polygons as consisting
of four straight lines (one on each side of the minimal bounding rectangle)
connected via directed walks, e.g. the left-most line is connected to the
top-most line via a walk taking steps only to the right and up and so on.
Mutually the directed walks have to respect the general restrictions for
a given class of polygons. We can get a contact where the
directed walks connect to the straight lines. From
this we see that a necessary  conditions for a CNAP is that
all straight pieces have length at least 2 (it is not sufficient since the
mutual neighbour avoidance of the directed walks is not automatically
guaranteed). The bijection is now obvious.
To go from a COP to a CNAP with 4 more steps we simply extend each straight
piece by a single step and to go from a CNAP to a COP we contract each
straight piece by a single step. That the first operation takes us from a
COP to a CNAP is clear. The convexity is clearly preserved and the extension 
of the straight pieces ensures that all occupied vertices are separated by 
an empty vertex and thus the condition on a NAP is fulfilled. That the reverse 
operation takes us from a CNAP to a COP is perhaps not as obvious. Note however 
that due to the neighbour avoidance there is at least one empty vertex between
any two vertices of the CNAP. So deletion operation can never result in overlapping
or crossing of occupied edges, at most we can get an osculation.

\subsection{Almost-convex polygons}

The generating function, $\ACPgf (x)$ for almost convex polygons was calculated 
exactly by Lin \cite{Lin92}, who found

\begin{equation}\label{eq:acp}
\ACPgf (x) = 16x^3A/[(1-x)(1-4x)^{5/2}]+4x^3B/[(1-x)(1-3x+x^2)(1-4x)^3],
\end{equation}
\noindent
where $A$ and $B$ are low-order polynomials with integer coefficients:

\begin{eqnarray*}
A & = & 1-9x+25x^2-23x^3+3x^4, \\
B & = & -4+56x-300x^2+773x^3-973x^4+535x^5-90x^6+24x^7.
\end{eqnarray*}

Given the close similarity between OP, NAP and SAP it is natural to look
for a solution of a form similar to Eq.~(\ref{eq:acp}). From the solution
to the convex case Eq.~(\ref{eq:copexp}) one would expect extra factors
of $(2+x)$ to turn up in the denominators. Using  standard mathematical
software packages it is easy to check for solutions of a form similar to
to Eq.~(\ref{eq:acp}). In this case we simply make an educated guess
for the denominator $D$ associated with $B$, multiply our series for  
almost-convex polygons with $D$ and look for a solution of the
form $A+B\sqrt{1-4x}$ by formally expanding this expression
and solving for the unknown coefficients in the polynomials $A$ and $B$.

In this manner we found the generating function for almost-convex
OP to be

\begin{equation}\label{eq:acop}
\ACOPgf (x) = 4x^2A^{\rm O}/[(2\! +\! x)(1\! -\! x)(1\! -\! 4x)^{5/2}]
 +4x^2B^{\rm O}/[(2\! +\! x)^2(1\! -\! x)^2(1\! -\! 3x\! +\!x^2)(1\! -\!4x)^3],
\end{equation}
\noindent
where:

\begin{eqnarray*}
A^{\rm O} & = & 2\! -\! 10x\! -\! 10x^2\! +\! 55x^3\! +\! 25x^4\! 
-\! 17x^5\! +\! 24x^6\! +\! 12x^7, \\
B^{\rm O} & = & \! -\! 2\! +\! 22x\! -\! 70x^2\! -\! 5x^3\! +\! 371x^4\! 
-\! 521x^5\! +\! 175x^6\! -\! 44x^7\! +\! 101x^8
                \! -\! 219x^9\! -\! 238x^{10}\! -\! 56x^{11}.
\end{eqnarray*}

Similarly we found the generating function for almost-convex NAP

\begin{equation}\label{eq:acnap}
\ACNAPgf (x) = 4x^4A^{\rm NA}/[(2\! +\! x)^2(1\! -\! x)^2(1\! -\! 4x)^{5/2}]
+4x^4B^{\rm NA}/[(2\! +\! x)^2(1\! -\! x)^2(1\! -\! 3x\! +\!x^2)(1\! -\!4x)^3],
\end{equation}
\noindent
where:

\begin{eqnarray*}
A^{\rm NA} & = & 2\! -\! 16x\! +\! 22x^2\! +\! 79x^3\! -\! 149x^4\! 
-\! 86x^5\! +\! 89x^6\! +\! 36x^7\! -\! 38x^8\! -\! 20x^9, \\
B^{\rm NA} & = & \!\! -\! 2\! +\! 26x\!\! -\!\! 112x^2\!\! +\!\! 111x^3\!\! 
+\!\! 434x^4\!\! -\!\! 1050x^5\!\! +\!\! 230x^6\!\! +\!\! 823x^7\!\! 
-\!\! 241x^8\!\! -\!\! 9x^9\!\! +\!\! 13x^{10}\!\! -\!\! 118x^{11}\!\! 
-\!\! 24x^{12}.
\end{eqnarray*}

\section{Conclusion}

We have derived long series for the generating function, radius of gyration
and area-weighted moments of osculating and neighbour-avoiding polygons.
Our extended series enables us to give precise estimate of the connective
constants $\mu^{\rm O}=2.677924128(5)$ for OP and $\mu^{\rm NA}=2.315845(4)$
for NAP. As expected these values are, respectively, larger and smaller
than the connective constant for SAP. The exponent estimates are consistent 
with the exact values $\alpha = 1/2$ and $\nu = 3/4$. We also obtain precise
estimates for the leading amplitudes. Analysis of the coefficients
of the radius of gyration series yielded results fully compatible with the 
prediction $BD=5/16\pi^2$ and from the first area-weighted moment we 
confirmed the universality of the amplitude ratio $E^{(1)}/D$.

In addition we used the series data to obtain the generating functions
for convex and almost-convex OP and NAP. In all cases these are very similar
to the corresponding generating functions for SAP.

\section*{E-mail or WWW retrieval of series}

The series for the various generating functions studied in this paper
can be obtained via e-mail by sending a request to 
I.Jensen@ms.unimelb.edu.au or via the world wide web on the URL
http://www.ms.unimelb.edu.au/\~{ }iwan/ by following the instructions.

\section{Acknowledgements}

This work was supported by a Fellowship and grant from the Australian Research
Council. Part of the calculations were performed on the facilities of
VPAC and APAC.

\clearpage

\clearpage

\begin{table}
\caption{\label{tab:OPupdate} 
The updating rules for the transfer-matrix algorithm for osculating
polygons. The new partial generating function is multiplied by a factor
$x^k$, where $k$ is the number of '1' and '2' entries in the local output
state. States marked by over-lining indicates instances in which two
loop-ends are joined and other edges have to be relabelled as described
by Enting \cite{Ent}.}
\begin{center}
\begin{tabular}{l|ccc} \hline \hline
  &   0   &    1    &   2  \\  \hline
0 & 00 12 &  01 10 & 02 20 \\
1 & 01 10 & $\overline{00}$ $\overline{12}$ 11 &  acc  12 \\
2 & 02 20 & 00 12 21 & $\overline{00}$ $\overline{12}$ 22 \\
\hline \hline
\end{tabular}
\end{center}
\end{table}

\begin{table}
\caption{\label{tab:NAPupdate} 
The updating rules for the transfer-matrix algorithm for neighbour avoiding
polygons. Similar remarks as for OP apply. States marked with a * are only
created conditionally depending on the states of nearby edges and states 
marked -- should never occur provided the conditional updating is implemented 
correctly.}
\begin{center}
\begin{tabular}{l|cccc} \hline \hline
  &    0      &       1         &        2        &  3 \\  \hline
0 & 00 $12^*$ &  $13^*$ $31^*$  &    $23^*$ 32    &  00 \\
1 & 13  31    & $\overline{33}$ &       add       &  -- \\
2 & $23^*$ 32 &      33         & $\overline{33}$ &  -- \\
3 &    00     &      --         &        --       &  00 \\
\hline \hline
\end{tabular}
\end{center}
\end{table}

\begin{table}
\caption{\label{tab:OPcrit} Estimates for the critical point
$x_c$ and exponents obtained from second order differential approximants to 
the series for generating function, the radius of gyration and first moments 
of area of square lattice osculating polygons. 
$L$ is the order of the inhomogeneous polynomial.}
\scriptsize
\begin{center}
\begin{tabular}{lllllll} \hline \hline
Series:  & \multicolumn{2}{c}{$\AM (u)$} 
         & \multicolumn{2}{c}{$\RG^2_g (u)$} 
         & \multicolumn{2}{c}{$\AM^{(1)} (u)$} \\ \hline 
$L$ & \multicolumn{1}{c}{$x_c$} & \multicolumn{1}{c}{$\alpha$} &
\multicolumn{1}{c}{$x_c$} & \multicolumn{1}{c}{$-(\alpha+2\nu)$} & 
\multicolumn{1}{c}{$x_c$} & \multicolumn{1}{c}{$2-(\alpha+2\nu)$} \\ \hline 
\multicolumn{7}{c}{Second order differential approximants}  \\ \hline
0  & 0.1394451590(29) & 1.5000193(66) & 
     0.139445388(43)  & -2.001275(91) & 
     0.139445489(67)  & -0.00113(18)  \\
2  & 0.1394451653(12) & 1.5000025(36) & 
     0.139445384(53)  & -2.00125(12)  & 
     0.139445339(72)  & -0.00084(16)  \\
4  & 0.1394451636(10) & 1.5000073(29) & 
     0.139445401(58)  & -2.00128(13)  & 
     0.139445255(37)  & -0.00061(46)  \\
6  & 0.1394451625(30) & 1.500012(12)  & 
     0.139445400(51)  & -2.00129(10)  & 
     0.13944532(10)   & -0.00075(27)  \\
8  & 0.1394451629(25) & 1.5000093(69) & 
     0.139445391(46)  & -2.001271(99) & 
     0.139445276(82)  & -0.00061(29) \\
10 & 0.1394451631(17) & 1.5000092(52) & 
     0.139445394(48)  & -2.00127(10)  &
     0.139445272(26)  & -0.000642(98) \\ \hline
\multicolumn{7}{c}{Third order differential approximants} \\ \hline
0  & 0.1394451623(29) & 1.5000122(92) &
     0.139445384(71)  & -2.00123(20)  & 
     0.13944529(10)   & -0.00064(30)  \\
2  & 0.1394451641(30) & 1.5000061(78) &
     0.139445444(54)  & -2.00138(12)  & 
     0.139445240(22)  & -0.000485(63) \\
4  & 0.1394451621(24) & 1.500014(10)  &
     0.139445448(53)  & -2.001381(78) & 
     0.139445209(68)  & -0.00036(25)  \\
6  & 0.1394451628(20) & 1.5000100(54) &
     0.139445397(69)  & -2.00126(19)  & 
     0.139445252(39)  & -0.00053(13)  \\
8  & 0.1394451637(10) & 1.5000078(37) &
     0.139445433(18)  & -2.001367(36) &
     0.139445212(50)  & -0.00037(21)  \\ 
10 & 0.1394451640(18) & 1.5000062(60) &
     0.13944541(13)   & -2.00118(45)  &
     0.139445272(83)  & -0.00056(23)  \\
\hline \hline
\end{tabular}
\end{center}
\end{table}

\begin{table}
\caption{\label{tab:NAPcrit} Estimates for the critical point
$x_c$ and exponents obtained from second order differential approximants to 
the series for generating function, the radius of gyration and first moments 
of area of square lattice neighbour-avoiding polygons. 
$L$ is the order of the inhomogeneous polynomial.}
\scriptsize
\begin{center}
\begin{tabular}{lllllll} \hline \hline
Series:  & \multicolumn{2}{c}{$\AM (u)$} 
         & \multicolumn{2}{c}{$\RG^2_g (u)$} 
         & \multicolumn{2}{c}{$\AM^{(1)} (u)$} \\ \hline 
$L$ & \multicolumn{1}{c}{$x_c$} & \multicolumn{1}{c}{$\alpha$} &
\multicolumn{1}{c}{$x_c$} & \multicolumn{1}{c}{$-(\alpha+2\nu)$} & 
\multicolumn{1}{c}{$x_c$} & \multicolumn{1}{c}{$2-(\alpha+2\nu)$} \\ \hline 
\multicolumn{7}{c}{Second order differential approximants}  \\ \hline
0  & 0.1864560(16) & 1.5026(17)  & 
     0.186403(12)  & -1.9453(65) &
     0.186424(31)  & 0.037(28)   \\
2  & 0.1864551(25) & 1.5033(30)  &
     0.186429(11)  & -1.9602(88) &
     0.1864488(42) & 0.0130(48) \\
4  & 0.1864532(56) & 1.5043(68)  &
     0.186433(14)  & -1.962(11)  &
     0.1864485(74) & 0.0133(84)  \\
6  & 0.1864538(16) & 1.5049(17)  &
     0.186425(11)  & -1.9566(70) &
     0.186448(12)  & 0.011(14)   \\
8  & 0.1864555(29) & 1.5030(32)  &
     0.186426(13)  & -1.9581(96) &
     0.186445(11)  & 0.016(11)   \\
10 & 0.1864527(63) & 1.5046(64)  &
     0.1864312(84) & -1.9605(70) &
     0.1864533(64) & 0.0070(74)  \\ \hline
\multicolumn{7}{c}{Third order differential approximants}  \\ \hline
0  & 0.1864567(30) & 1.5011(39)  &
     0.186442(28)  & -1.983(32)  &
     0.186446(10)  & 0.0155(99)  \\
2  & 0.1864572(13) & 1.5011(16)  &
     0.186416(43)  & -1.960(39)  &
     0.1864505(56) & 0.0113(56)  \\
4  & 0.1864573(22) & 1.5007(30)  & 
     0.186447(20)  & -1.975(47)  &
     0.186449(11)  & 0.010(13)   \\
6  & 0.1864563(22) & 1.5020(27)  &
     0.1864493(79) & -1.9872(96) &
     0.1864513(44) & 0.0097(45)  \\
8  & 0.1864568(40) & 1.5006(60)  &
     0.186450(13)  & -1.988(18)  &
     0.186437(25)  & 0.022(29)   \\
10 & 0.1864550(37) & 1.5037(44)  &
     0.186461(19)  & -2.005(28)  &
     0.186439(24)  & 0.024(32)   \\ 
\hline \hline
\end{tabular}
\end{center}
\end{table}

\end{document}